\documentclass[final]{svjour3}
\usepackage[dvipdfmx]{graphicx}
\usepackage{caption}
\usepackage{subcaption}
\usepackage{rotating}
\usepackage{amssymb}
\usepackage{mathptmx}
\usepackage[numbers]{natbib}
\usepackage{xcolor}
\makeatletter
\journalname{Journal of Low Temperature Physics}
\newcommand{\apj}{The Astrophysical Journal}

\bibpunct{[}{]}{,}{n}{}{,}

\begin{document}

\newcommand{\hdblarrow}{H\makebox[0.9ex][l]{$\downdownarrows$}-}

\title{Testing low-loss microstrip materials with MKIDs for microwave applications}
\def\Vanderbilt{a}
\def\ANLHEP{b}
\def\KICPChicago{c}
\def\EFIChicago{d}
\def\PhysicsUChicago{e}
\def\AAUChicago{f}
\def\supit#1{\raisebox{0.8ex}{\small\it #1}\hspace{0.05em}}

\author{
   J. Hood\protect\supit{\Vanderbilt,\KICPChicago,\AAUChicago} \and
   P.~S. ~Barry\protect\supit{\ANLHEP,\KICPChicago} \and
   T.~Cecil\protect\supit{\ANLHEP} \and
   C.~L.~Chang\protect\supit{\ANLHEP,\KICPChicago,\AAUChicago} \and
   J.~Li\protect\supit{\ANLHEP} \and
   S.~S.~Meyer\protect\supit{\KICPChicago,\EFIChicago,\PhysicsUChicago,\AAUChicago} \and
   Z.~Pan\protect\supit{\KICPChicago,\PhysicsUChicago} \and
   E.~Shirokoff\protect\supit{\KICPChicago,\AAUChicago} \and
   A.~Tang\protect\supit{\KICPChicago,\AAUChicago} \and
}

\institute{
   \protect\supit{\Vanderbilt}Vanderbilt University,Nashville, TN 37235 \and
   \protect\supit{\KICPChicago}Kavli Institute for Cosmological Physics, Univ. of Chicago, 5640 S. Ellis Ave., Chicago, IL 60637 \and
   \protect\supit{\ANLHEP}Argonne National Laboratory, High-Energy Physics Division, 9700 S. Cass Ave., Argonne, IL 60439 \and
   \protect\supit{\AAUChicago}Dept. of Astronomy and Astrophysics, Univ. of Chicago, 5640 S. Ellis Ave., Chicago, IL 60637 \and
   \protect\supit{\EFIChicago}Enrico Fermi Institute, Univ. of Chicago, 5640 S. Ellis Ave., Chicago, IL 60637 \and
   \protect\supit{\PhysicsUChicago}Dept. of Physics, Univ. of Chicago, 5640 S. Ellis Ave., Chicago, IL 60637 \and
   \protect\supit{\AAUChicago}Univ. of Chicago, 5640 S. Ellis Ave., Chicago, IL 60637 
}

\maketitle

\begin{abstract}

Future measurements of the millimeter-wavelength sky require a low-loss super- conducting microstrip, typically made from niobium and silicon-nitride, coupling the antenna to detectors. We propose a simple device for characterizing these low-loss microstrips at 150 GHz. In our device we illuminate an antenna with a thermal source and compare the measured power at 150 GHz transmitted down microstrips of different lengths. The power measurement is made using Microwave Kinetic Inductance Detectors (MKIDs) fabricated directly onto the microstrip dielectric, and comparing the measured response provides a direct measurement of the microstrip loss. Our proposed structure provides a simple device (4 layers and a DRIE etch) for characterizing the dielectric loss of various microstrip materials and substrates. We present initial results using these devices. We demonstrate that the millimeter wavelength loss of microstrip lines, a few tens of millimeters long, can be measured using a practical aluminum MKID with a black body source at a few tens of Kelvin.


\end{abstract}

\section{Introduction}

Superconducting microstrip is commonly used in detector arrays for measuring the Cosmic Microwave Background (CMB) where it serves as the transmission line for the optical signal. The CMB is an imprint of the universe from when it was only $\sim$380\,K years old and provides unique insight into fundamental physics. Current and upcoming measurements of the CMB temperature and polarization require large arrays of detectors observing from $\sim$30\,GHz to 300\,GHz, and superconducting microstrip is a key technology for all state-of-the-art CMB architectures as seen in ~\citep{BICEP22015,Suzuki2012,Posada2018, Duff2016, Rostem2012, Abazajian2016}. 
In this paper, we introduce a device that uses Microwave Kinetic Inductance Detectors to measure the microstrip dielectric loss at both a few GHz and $\sim$220\,GHz. We present our device design and fabrication, a measurement of the low frequency loss for our materials, and the predicted outcome of our measurement at $\sim$220\,GHz.

\section{Design}
\begin{figure}[t]
\begin{center}
\includegraphics[width=0.4\linewidth, keepaspectratio]{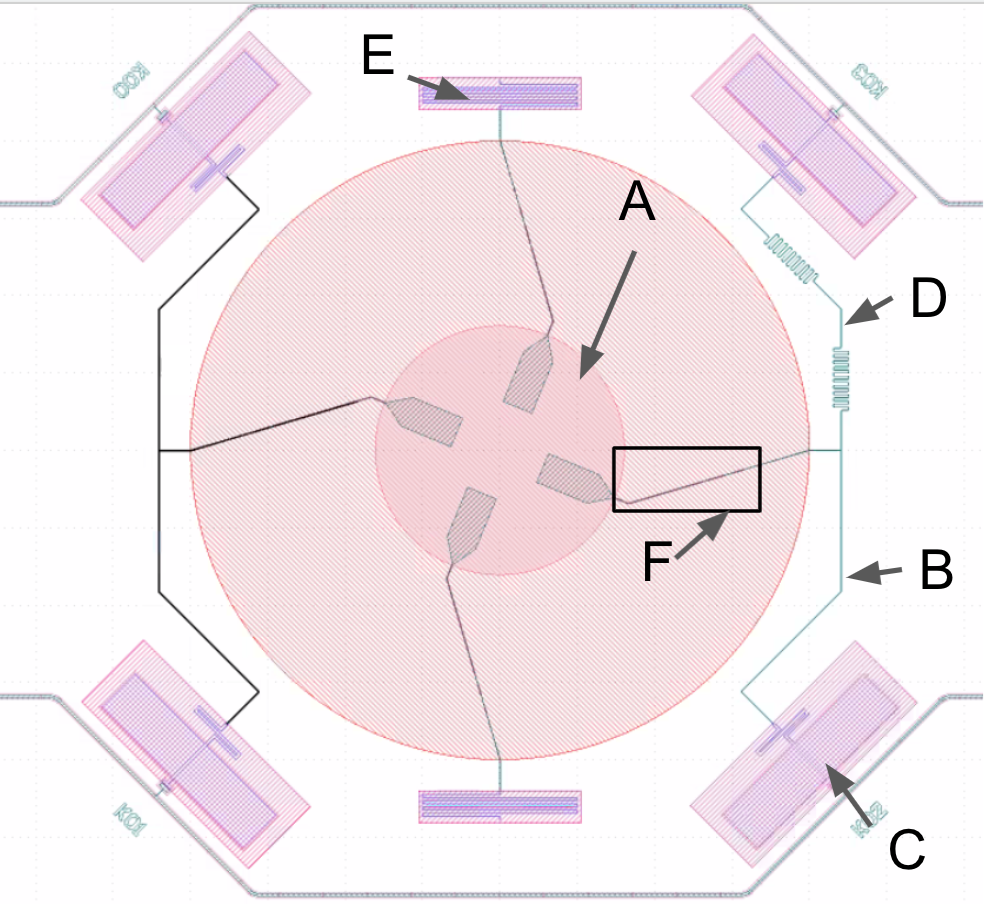}
\caption{\label{fig:pixel} OMT with antennae probes (A),
short microstrip (B) to MKID detector (C) and long microstrip (D). The long microstrip length varies across multiple devices ranging from 2.65\,mm to 200\,mm longer in length. Terminating resistor (E). F is the CPW to microstrip transition.
}
\end{center}
\end{figure}

Our design is shown in Fig.~\ref{fig:pixel}. Incoming light feeds our detectors via feed horns and then couples to our device through a planar ortho-mode transducer (OMT). From the OMT probes the signal travels through a Co-Planar Waveguide (CPW)-to-microstrip transition consisting of alternating sections of CPW and microstrip waveguide. The microstrip consists of a ground and top conductor layer of niobium (Nb) with a dielectric layer of silicon nitride. In each pixel, each polarization couples to two pairs of detectors where the power is split in half and transferred to the two detectors through different lengths of transmission lines. Each of the detectors are Al MKID designed with parallel plate capacitors so that they are also sensitive to the underlying dielectric and can then measure the loss at a few GHz.

\section{Device Fabrication}

\begin{figure}[t]
\begin{center}
\includegraphics[width=0.6\linewidth, keepaspectratio]{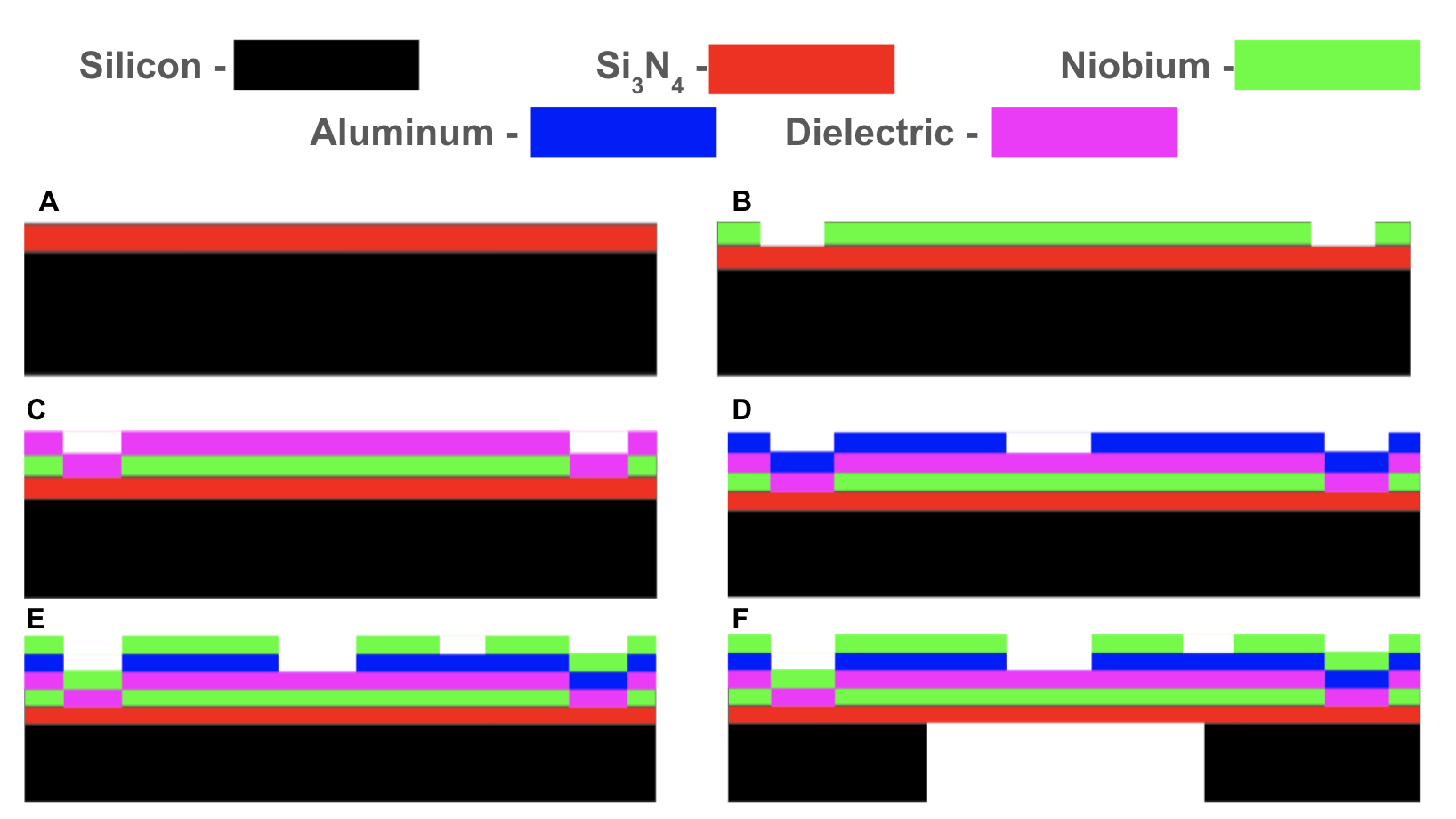}
\caption{
Outline of the fabrication process. The side profile shows layers of Si (black), Si$_{3}$N$_4$ (red), Nb (green), Al (blue), dielectric (pink). A) Si wafer with Si$_{3}$N$_4$ layer deposited. B) First layer of Nb. C) Dielectric layer. D) Al layer. E) Top Nb layer. F) Handle wafer etch done to open OMT window.
}
\label{fig:fig2}
\end{center}
\end{figure}

Fabrication is done on a double-sided polished high resistivity 500\,$\mu$m thick silicon (Si) wafer with one side coated by 450\,nm of thermally grown SiO$_{2}$ and 2\,$\mu$m of Low Pressure Chemical Vapor Deposition (LPCVD) Si$_{3}$N$_4$. These layers were both commercially deposited, the commercial Si$_{3}$N$_4$ layer serves as the structural support for the OMTs. The 450\,nm SiO$_{2}$ layer is the etch stop layer for the deep silicon etching process Fig.\ref{fig:fig2}. The first layer of commercially deposited silicon nitride (Si$_{3}$N$_4$), step A (red) acts as a stop layer for the deep silicon etch done in the last step. Next, we deposit 300\,nm of Nb using sputtering, which is then patterned and etched using inductively coupled plasma (ICP) etching (step B) to make the ground plane. We then grow a layer of SiN$_3$ with a thickness of 500\,nm, which makes up the dielectric layer of our microstrip (step C). Next, a 30\,nm thick layer of aluminum (Al) is subsequently deposited using sputtering, patterned, and wet etched using a standard phosphoric acid-based Al etchant (step D), which forms the Al capacitor pads and Al inductors of our MKIDs. The last metal layer consists of a 300\,nm layer of Nb, which is deposited via sputtering, patterned, and plasma etched (step E). This layer makes up the Nb wiring of our device, including the CPW readout lines, OMT probes, and the CPW part of our CPW-to-microstrip transition section.

In the last step (step F), we use a deep silicon etch to remove the handle wafer from underneath the OMT and capacitor part of our MKIDs. This is done by first spinning a protective resist layer on the front surface of the wafer. We then flip the wafer, pattern the backside and use deep reactive ion etching (DRIE) to etch until the Si$_{3}$N$_4$ stop layer. Hydrofluoric acid (HF) bath is then used to remove the SiO$_{2}$ stop layer, which would otherwise contribute significant two-level system (TLS) loss and noise to our MKID. We also release the chip itself in this final step. The outer dimensions of the chip are defined during the deep silicon etch. We leave several tabs at the corners of the chip to physically connect it to the wafer frame during subsequent processing. After all processing is complete the chip is snapped from the wafer frame, breaking the tabs. 

\section{Testing Results}
\begin{figure}[htbp]
\begin{center}
\includegraphics[width=0.6\linewidth, keepaspectratio]{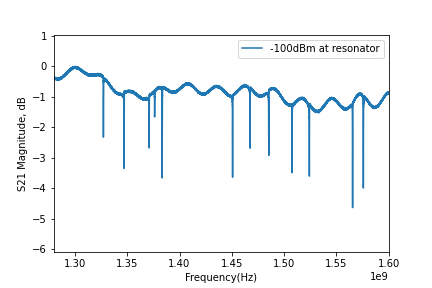}
\caption{\label{fig:sweeps}
S21 vs. frequency sweep of the mm-loss device. All 12 designed resonators appear as dips with 100\% yield. 
}
\end{center}
\end{figure}

We fabricated our first batch of test devices using aluminum for both the capacitor and inductor. In Fig.~\ref{fig:sweeps} we show the S21 vs. frequency sweep measured with a vector network analyzer (VNA). There are 12 resonators in total for a yield of 100\%. The resonance dip depth changes as a function of driving power at the resonator, and the bifurcation power for these resonators is around -85\,dBm. 

We measure temperature-dependent frequency shifts at high power levels as probes for dielectric loss at low frequencies below 10~GHz as described in \citep{Carter2019}. Here we measure the S21 of each resonator as a function of frequency in a small segment near the resonant frequency and fit for the resonance frequency. We then vary the stage temperature and plot the resonance frequency as a function of operating temperature. The frequency shift at lower temperatures can be described by a two-level system (TLS) model which describes the tunneling states in amorphous solids and their coupling to external electrical fields. The frequency shift is described by \citep{Phillips1987, Pappas2011}:

\begin{figure}[h]

\begin{subfigure}{0.5\textwidth}
\includegraphics[width=1.0\linewidth, height=6cm]{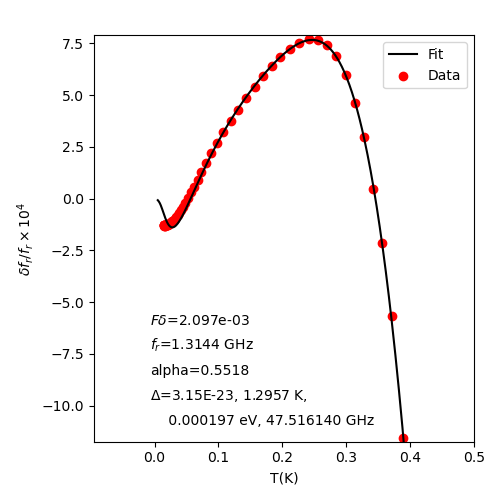} 
\caption{Frequency vs. temperature}
\label{fig:tls}
\end{subfigure}
\begin{subfigure}{0.5\textwidth}
\includegraphics[width=0.99\linewidth, height=6cm]{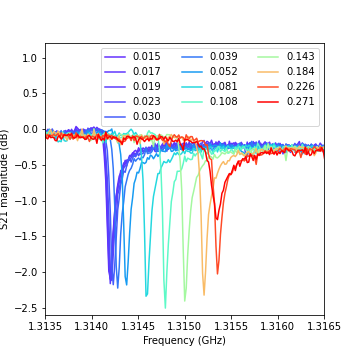}
\caption{Temperature sweep}
\label{fig:subim2}
\end{subfigure}

\caption{Left: the resonator fractional frequency shift as a function of temperature and the fit to Eqn. \ref{eq:tls} as well as the higher-temperature model described in \citep{Gao2008}. This shows the measured loss at $\sim$1-10\,GHz. Right: resonance S21 as a function of temperature. We extracted the resonance frequencies from the S21 measurements and fit frequency as a function of temperature.}
\label{fig:tls_whole}
\end{figure}

\begin{equation}
\frac{f(T)-f_{0}}{f_{0}}=\frac{F \delta_{\mathrm{TLS}}^{0}}{\pi}\left[{\mathrm{Re\Psi}}\left(\frac{1}{2}-\frac{\hbar f_{0}}{j k_{B} T}\right)-\log \frac{\hbar f_{0}}{k_{B} T}\right]
\label{eq:tls}
\end{equation}
Where $\mathrm{\Psi}$ is the complex digamma function, $F$ is the filling factor of the TLS, $\delta_{\mathrm{TLS}}^{0}$ is the loss tangent, $f(T)$ is the temperature-dependent resonant frequency, and $f_0$ is the resonator frequency at zero temperature. At higher temperatures, the frequency shift is dominated by the breaking of Cooper pairs, and the model is discussed in \citep{Gao2008}. We fit the data to the model and obtained $F\delta_{\mathrm{TLS}}^{0}=2\times10^{-3}$ for the resonators. The electrical field is mostly confined in the dielectric for our design, so $F$ is close to one, and this indicates our loss tangent $\delta_{\mathrm{TLS}}^{0}$ is close to $2\times10^{-3}$, which is typical for silicon nitride, the material we used for our dielectric. Note that the measurements done above are with a sample enclosed by an aluminum box, and this sample is not coupled to an optical source. The next step is to fabricate devices with aluminum inductors and Nb capacitors and couple the devices to an optical load.

\section{Forecast for Microstrip-Loss Measurement at $\sim$ 220\,GHz}
In the previous section, we demonstrated that we could fabricate these devices with a 100\% yield and measure the dielectric loss tangent at $\sim$10\,GHz. We can also measure the loss at $\sim$220\,GHz by illuminating our device with a black body and comparing the response between detectors with different lengths of transmission lines. Here, we predict the MKID optical response for varying microstrip lengths and a black body temperature of 40\,K. To model this, we first calculate the expected power that will be transmitted down the different microstrip lengths and absorbed by our detectors. Second, we calculate the number of quasiparticles generated by this absorbed power. Finally, we use our measured frequency shift as a function of temperature to convert the number of quasiparticles into the predicted detector response.

To calculate the optical power absorbed by each detector, we use the Rayleigh Jeans approximation and the following equation:
\begin{equation}
P = k_{b}*T*\Delta\nu*\eta_{f}^2e^{\frac{-(d*L1)}{\lambda}}0.5e^{\frac{-(d*L2)}{\lambda}},
\label{eq:pow}
\end{equation}
where: $k_{b}$ is the Boltzman constant, T is the black body temperature, $\Delta\nu$ the optical bandwidth (which we take as 25\,GHz), $\eta_{f}$ is the transmission for each of our two free space filters (which we take as 0.95), $\lambda=7.559\times10^{-4}$ is the wavelength in microstrip, and L1 is the length of microstrip from the OMT paddles to the power splitter (which is $\sim$1.1\,mm). There is a factor of 0.5 to account for the power splitting between the two detector arms and L2 is the varying length of microstrips (2.65\,mm, 10\,mm, 20\,mm, 30\,mm and 200\,mm) and d is the dielectric loss tangent. To determine the number of quasiparticles from the absorbed optical power, we use $N_{qp}=P\tau_{qp}$/$\eta\Delta$. Since $\tau _{qp}$ $\propto$ $N_{qp}^{-1}$ as seen in \citep{Baselmans2012}, we are left with the following equation:

\begin{equation}
N_{qp}(P)=\sqrt{\frac{P\tau_{qp,{0.25K}}N_{qp,{0.25K}}}{\eta\Delta}},
\label{eq:nqpp}
\end{equation}
where $\tau_{qp,{0.25K}} = 0.05$\,ms is the quasiparticle lifetime we measure from the our Generation Recombination (GR) noise roll-off at 250\,mK, and $N_{qp,{0.25K}} = 595.3$ is the calculated quasiparticle number at 250\,mK.

\begin{figure}[htbp]
\begin{center}
\includegraphics[width=0.6\linewidth, keepaspectratio]{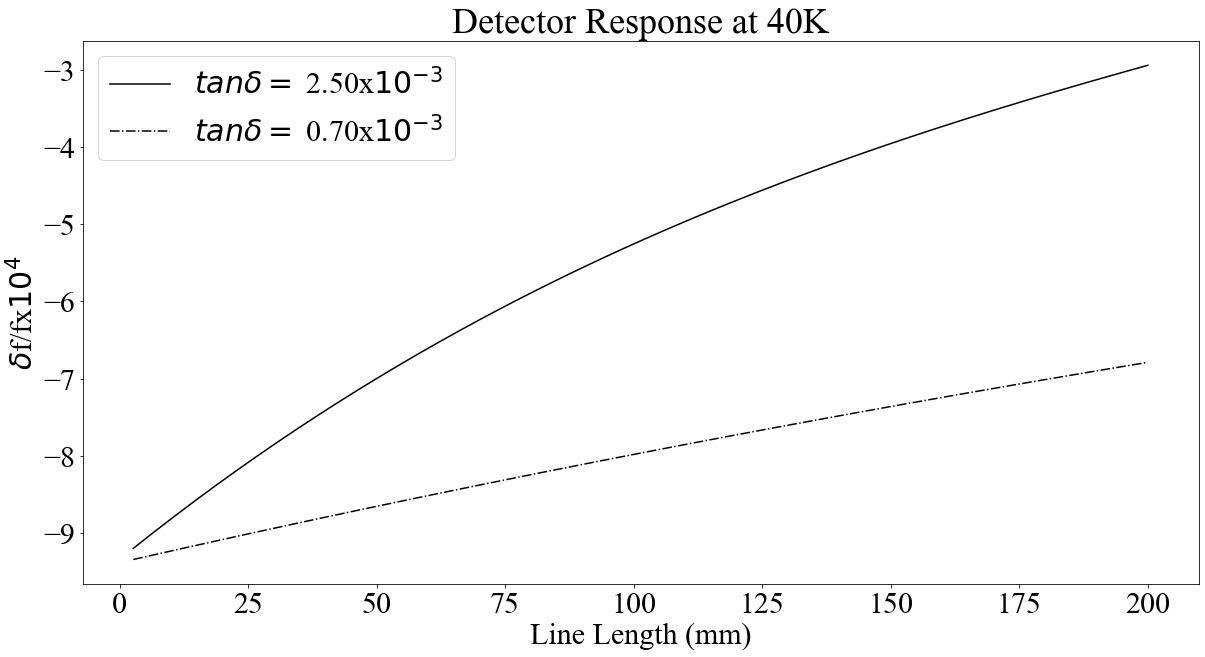}
\caption{
Predicted fractional frequency shift as a function of differential microstrip length at $\sim$220\,GHz. Both lines are coupled to the same black body with a temperature of 40\,K. 
}
\label{fig:df}
\end{center}
\end{figure}

To calculate the change in detector resonance frequency due to the change in quasiparticle number, we use the slope of the high temperature data shown in Fig.~\ref{fig:tls}. Fig.~\ref{fig:df} shows our prediction for the detector fractional frequency shift ($\delta$f/f) for the various microstrip lengths, a black body temperature of 40\,K, and two different loss tangents. The choice of loss tangents is motivated by the measurements in the previous section and \citep{Duff2016}. Since we can typically measure $\delta$f/f down to values of $10^{-7}$ we expect to have no issue measuring $\delta$f/f at a level of $10^{-4}$.
 
\section{Summary}
In conclusion, for the loss tangents that we expect these materials to have, at reasonable black body source temperatures and microstrip lengths, our predicted $\delta$f/f is large enough for us to measure the loss tangent. Therefore we have demonstrated that the mm-wavelength loss of microstrip lines, only a few tens of mm long, can be measured using a practical aluminum MKID with a black body source at a few tens of Kelvin.

\begin{acknowledgements}
Work at Argonne, including use of the Center for Nanoscale Materials, an Office of Science user facility, was supported by the U.S. Department of Energy, Office of Science, Office of Basic Energy Sciences and Office of High Energy Physics, under Contract No. DE-AC02-06CH11357, and the SCGSR Fellowship.
\end{acknowledgements}

\end{document}